\documentclass[twocolumn]{aastex631}

\usepackage{graphicx}
\usepackage{amsmath}
\usepackage{amssymb}
\usepackage{booktabs}
\usepackage{xcolor}
\definecolor{plum}{HTML}{5C2569}
\definecolor{orchid}{HTML}{8B3E8F}
\definecolor{rose}{HTML}{B85C9E}
\hypersetup{colorlinks=true, citecolor=plum, linkcolor=orchid, urlcolor=rose}

\newcommand{\flare}{\textsc{flare}}
\newcommand{\luna}{\textsc{luna}}
\newcommand{\astranet}{\textsc{astra}Net}
\newcommand{\applecider}{AppleCiDEr}
\newcommand{\boom}{\textsc{boom}}
\newcommand{\alerce}{\textsc{alerce}}
\newcommand{\ztf}{ZTF}
\newcommand{\Mpseudo}{M_{\mathrm{pseudo}}}

\newcommand{\tpr}{\mathrm{TPR}}

\begin{document}

\title{Photometric classification and out-of-taxonomy detection of ZTF transients with FLARE}

\author[0000-0001-7357-0889]{Argyro Sasli}
\correspondingauthor{Argyro Sasli}
\email{asasli@umn.edu}
\affiliation{School of Physics and Astronomy, University of Minnesota, Minneapolis, MN 55455, USA}

\author[0000-0001-7129-1325]{Felipe Fontinele Nunes}
\affiliation{School of Physics and Astronomy, University of Minnesota, Minneapolis, MN 55455, USA}

\author{Xuanrui Zhang}
\affiliation{School of Statistics, University of Minnesota, Minneapolis, MN 55455, USA}
\author[0009-0007-2863-520X]{Dimitrios Kourtesis}
\affiliation{School of Physics, Faculty of Sciences, Aristotle University of Thessaloniki, 54124 Thessaloniki, Greece}
\author{Saina Nikmehr}
\affiliation{Cahill Center for Astrophysics, California Institute of Technology, MC 249-17, 1216 E California Boulevard, Pasadena, CA 91125, USA}

\author[0000-0002-9380-7983]{Alexandra Junell}
\affiliation{School of Physics and Astronomy, University of Minnesota, Minneapolis, MN 55455, USA}

\author[0000-0003-1546-6615]{Jesper Sollerman}
\affiliation{Department of Astronomy, Stockholm University, 10691 Stockholm, Sweden}

\author[0009-0009-7000-8343]{Antoine Le Calloch}
\affiliation{School of Physics and Astronomy, University of Minnesota, Minneapolis, MN 55455, USA}
\affiliation{NSF Institute on Accelerated AI Algorithms for Data-Driven Discovery (A3D3)}
\author[0000-0003-1314-4241]{Sushant Sharma Chaudhary}
\affiliation{School of Physics and Astronomy, University of Minnesota, Minneapolis, MN 55455, USA}
\author[0000-0001-7922-8494]{Sneha Maharjan}
\affiliation{School of Physics and Astronomy, University of Minnesota, Minneapolis, MN 55455, USA}

\author[0009-0002-5097-3188]{Maojie Xu}
\affiliation{Department of Computer Science \& Engineering, University of Minnesota, Minneapolis, MN 55455, USA}
\author[0009-0005-9422-0091]{Avyukt Raghuvanshi}
\affiliation{School of Physics and Astronomy, University of Minnesota, Minneapolis, MN 55455, USA}

\author[0000-0001-5668-3507]{Steven L. Groom}
\affiliation{IPAC, California Institute of Technology, 1200 E. California
             Blvd, Pasadena, CA 91125, USA}
\author[0000-0002-0987-3372]{Joahan Castaneda Jaimes}
\affiliation{Division of Physics, Mathematics and Astronomy, California
    Institute of Technology, 1200 E. California Blvd, Pasadena, CA 91125, USA}
\author[0000-0003-1227-3738]{Josiah Purdum}
\affiliation{Cahill Center for Astrophysics, California Institute of Technology, Pasadena, CA, 91125, USA}
\author[0000-0001-7062-9726]{Roger Smith}
\affiliation{Caltech Optical Observatories, California Institute of Technology, Pasadena, CA  91125}

\author[0000-0002-5619-4938]{Mansi M. Kasliwal}
\affiliation{Division of Physics, Mathematics, and Astronomy, California Institute of Technology, Pasadena, CA 91125, USA}
\author[0000-0002-3168-0139]{Matthew J. Graham}
\affiliation{Cahill Center for Astrophysics, California Institute of Technology, Pasadena, CA 91125, USA}
\author[0000-0003-2242-0244]{Ashish A. Mahabal}
\affiliation{Division of Physics, Mathematics and Astronomy, California Institute of Technology, Pasadena, CA 91125, USA}
\affiliation{Center for Data Driven Discovery, California Institute of Technology, Pasadena, CA 91125, USA}
\author[0000-0002-8262-2924]{Michael Coughlin}
\affiliation{School of Physics and Astronomy, University of Minnesota, Minneapolis, MN 55455, USA}

\begin{abstract}
Photometric classifiers decide which of the million nightly alerts of time-domain surveys receive one of the few thousand spectra taken each year, and they are the first place where a transient outside the trained classes can be noticed. We present \flare, the photometric classification and anomaly-detection module of \applecider, evaluated on $10\,497$ spectroscopically confirmed \ztf\ transients in six classes and $53$ objects from families outside the taxonomy. Catalogue context raises the macro $F_1$ score, with the largest gains where the defining information is absent from the light curve. Blind detection recovers few out-of-taxonomy objects at a $1\%$ false-alarm rate, while a simulated review loop recovers $83\%$ of them after inspecting $15\%$ of the stream. The first confirmations of a new family are most useful in the feature space: with the same eleven labels, a head on the frozen features detects $68\%$ of the tidal disruption events hidden from the training against $57\%$ for a combiner over the anomaly scores, and retraining with them as a new class reaches $85\%$. The module runs in production on Fritz, the \ztf\ instance of SkyPortal, and its alarm is reported as a false-alarm rate per unit time on the stream, calibrated on the photometry that the service receives.
\end{abstract}


\section{Introduction}
\label{sec:intro}

Since the start of its operations in 2018, the Zwicky Transient Facility (\ztf; \citealp{bellm2019ztf, graham2019ztf, 2020PASP..132c8001D}) has distributed up to about a million alerts per night \citep{patterson2019ztf}, and the Vera C. Rubin Observatory \citep{ivezic2019lsst}, which began distributing alerts in February 2026, is expected to reach about ten million per night once the survey is fully operational. Only a few thousand of these transients are classified spectroscopically each year, so that photometric classifiers decide, in practice, which alerts receive a spectrum, and they are also the first place where a new kind of transient can be noticed, as an object that matches none of the trained classes.

So far, several photometric classifiers have been developed for this task \citep{muthukrishna2019rapid, forster2021alerce, sanchezsaez2021alerce, villar2019}, and many of them run today inside the alert brokers \citep{smith2019lasair, moller2021fink, matheson2021antares, boom2025}. Anomaly searches have used light-curve features \citep{malanchev2021light}, expert feedback \citep{lochner2021astronomaly, ishida2021active} and the latent space of a trained classifier \citep{perezcarrasco2023mcdsvdd, gupta2025mcif}. These tools address successive stages of discovery, in which an unfamiliar object is flagged, confirmed by a human and finally represented by a trained class, and in this work we compare the three stages on the same objects, with the aim of measuring how the first confirmations of a new family should be used.

A second question concerns the information available to each stage. Some class distinctions depend on the luminosity or on the position within the host galaxy, which public catalogues can supply at the time of the alert, while others depend on spectral features. For this reason, we interpret the errors of the classifier and the recovery of the anomalies in terms of these observables, and we measure the contribution of each block of catalogue context; the models themselves are empirical, and the physics enters through the organisation of the analysis.

Here, we introduce \flare, the photometric module of the \applecider\ framework \citep{applecider2025}, which is built for the \boom\ alert broker \citep{boom2025} and whose spectroscopic module, \astranet, provides classifications, anomaly scores and prediction sets for spectra \citep{sasli2026astranet, applecider2025spectranet}. \flare\ uses the \ztf\ photometry and public catalogues, without spectra or image cutouts, and we evaluate it on $10\,497$ spectroscopically confirmed transients and on $53$ objects from families outside its six-class taxonomy.

We focus on three aspects of this workflow: (i) the information that the external context adds to the light curve (Section~\ref{sec:classification}); (ii) the recovery of the out-of-taxonomy objects by blind ranking, active review and learning from the first labels (Sections~\ref{sec:anomaly} and~\ref{sec:lifecycle}); and (iii) the reliability of both layers under missing context, short observation horizons and the transfer to the sources of Fritz (Sections~\ref{sec:pitfalls}--\ref{sec:deployment}). Finally, we discuss the limitations in Section~\ref{sec:discussion} and present our conclusions in Section~\ref{sec:conclusions}.

\section{Benchmark}
\label{sec:data}

\subsection{Spectroscopically confirmed sample}
The labels come from the \ztf\ Bright Transient Survey \citep[BTS;][]{fremling2020bts, perley2020bts}, a magnitude-limited catalogue in which $12\,916$ of $20\,547$ objects have a spectroscopic type; we impose no cut beyond its peak limit of $19$. The classified fraction is $81\%$ for objects brighter than $18.5$ and $41\%$ for fainter ones, so that the $7\,631$ unclassified objects are on average half a magnitude fainter (Section~\ref{sec:discussion}). The catalogue carries the selection function of the survey: BTS saves extragalactic candidates, so that the $304$ cataclysmic variables (CVs) and $251$ active galactic nuclei (AGN) of the benchmark are the members of these classes that most resembled supernovae, a harder set than the raw stream, and tidal disruption events (TDEs) enter by the same route, while dedicated photometric searches recover them more completely \citep{stein2024tdescore}.

We retrieve the \ztf\ photometry \citep{masci2019ztf} of each object through the \boom\ broker \citep{boom2025}, and we require at least eight detections, with at least two in each of the $g$ and $r$ bands, within a $100$-day window from the first detection; the $i$ band is used where present, but it is absent from the public alert stream and contributes $1.2\%$ of the detections, so that we describe the light curves as two-band. In total, $10\,497$ objects of the six trained classes and $52$ of families outside the taxonomy pass these criteria, and we add AT\,2018cow (ZTF18abcfcoo), the prototypical fast blue optical transient, which passes every cut but is listed as ``other'' in the catalogue; the out-of-taxonomy hold-out thus contains $53$ objects, and the benchmark $10\,550$.

\begin{table}[t]
\footnotesize\setlength{\tabcolsep}{2.5pt}
\caption{Composition of the benchmark: the six trained classes and the out-of-taxonomy hold-out. The sub-types included in each class are listed in Appendix~\ref{app:taxonomy}.}
\label{tab:data}
\begin{tabular}{lrl}
\hline\hline
class & $N$ & content \\
\hline
SN\,Ia   & 7159 & thermonuclear, incl.\ 91T/91bg/Iax/Ia-CSM \\
SN\,CC   & 2559 & core collapse: II, IIn, IIb, Ib/c, Ibn \\
CV       &  304 & cataclysmic variables (Galactic) \\
AGN      &  251 & active galactic nuclei \\
SLSN     &  156 & superluminous supernovae \\
TDE      &   68 & tidal disruption events \\
\hline
\multicolumn{3}{l}{\emph{out-of-taxonomy hold-out (never trained)}}\\
novae    &   37 & classical novae \\
novel    &   16 & LBV 6, LRN 4, ILRT 3, Ca-rich 2, FBOT 1 \\
\hline
total    & 10\,550 & $10\,497$ trained $+$ $53$ held out \\
\hline\hline
\end{tabular}
\end{table}

The composition of the benchmark is given in Table~\ref{tab:data}. The six trained classes are a coarsening of the taxonomy of the companion spectroscopic classifier \astranet\ \citep{sasli2026astranet}, in which its Ibc and II classes are merged into a single core-collapse class, so that the spectroscopic probabilities map onto the photometric classes by summation and the two predictions can be combined. The families of the hold-out play the role of the unknown classes in this work (Section~\ref{sec:ceiling}).

\subsection{What a light curve carries, class by class}
\label{sec:physics}

For the interpretation of the following sections, we summarise for each class the property that distinguishes it and whether a two-band light curve can measure it. The shape and the colour evolution identify most Type Ia supernovae, while the stripped-envelope core-collapse supernovae can resemble them at the \ztf\ cadence and are separated by spectral lines, which motivates the test of the photometric granularity in Section~\ref{sec:granularity}. Superluminous supernovae are defined by their luminosity \citep{galyam2019slsn}, so that a distance is needed to estimate it, although their duration and bumps carry some information \citep{russeil2026nomai}.

Tidal disruption events are blue, slowly fading nuclear transients \citep{gezari2021tde, vanvelzen2021tde, stein2024tdescore}, so that the position within the host separates them from the supernovae and a persistent nuclear source marks the AGN. The off-nuclear TDEs found recently \citep{somalwar2025offset, stein2026offnuclear} follow a different pattern, and either layer may miss them. Cataclysmic variables are Galactic binaries with recurrent outbursts, and the absence of a host identifies them, provided that it is distinguished from a host below the limit of the catalogue.

The hold-out contains LBV outbursts, which can resemble Type IIn supernovae \citep{vandyk2000impostors}, calcium-rich transients, which are identified by their late-time spectra \citep{deK2021carich}, LRNe and ILRTs, which overlap faint red supernovae, the rapidly evolving FBOT AT\,2018cow \citep{ho2023fbot}, and classical novae, i.e. thermonuclear outbursts in CV systems. These physical distinctions motivate the tests of the following sections, while the separability of each family is measured empirically.

\subsection{Host-galaxy context}
We use four external blocks, i.e. the Pan-STARRS host photometry and angular offset, a photometric redshift from the Legacy Surveys, the Gaia and WISE point-source context, and a Pan-STARRS pre-outburst counterpart, and we give their query definitions, match rates and catalogue limitations in Appendix~\ref{app:context}. The photometric redshift supplies a pseudo-absolute magnitude,
\begin{equation}
\Mpseudo = m_{\mathrm{peak}} - 5\log_{10}\!\left[d_L(z_{\mathrm{phot}})/10\,\mathrm{pc}\right],
\label{eq:mpseudo}
\end{equation}
where $d_L$ is the luminosity distance in a flat $\Lambda$CDM cosmology with $H_0=70\,\mathrm{km\,s^{-1}\,Mpc^{-1}}$ and $\Omega_m=0.3$. Spectroscopic redshifts enter only the reference comparison of Section~\ref{sec:distance}, because their availability depends strongly on the class (Section~\ref{sec:missingness}); at deployment, a known host redshift can replace the photometric one, a substitution that lies outside the benchmark evaluated here.

The host offset separates nuclear from displaced events, the WISE colour marks persistent AGN activity, and the pre-outburst counterpart reveals quiescent binaries and nuclei. The missing blocks are kept and handled by the context dropout of Section~\ref{sec:dropout}, and since a nearest-source association can select the wrong galaxy, a change of catalogue or of association rule requires the corresponding features to be refitted.

\section{Methods}
\label{sec:methods}

\subsection{Light-curve representation}
From each light curve, i.e. a set of irregularly sampled measurements $(t, \mathrm{flux}, \sigma_{\mathrm{flux}}, \mathrm{band})$, we compute $160$ features, which include the per-band statistics of the \texttt{light-curve} library \citep{malanchev2021light}, fits of the Bazin function \citep{bazin2009}, colours and their evolution, and descriptors of the cadence. Missing bands and failed fits are kept as NaN values, which gradient-boosted trees \citep{ke2017lightgbm} handle natively.

For the neural network that we use in the comparison of Section~\ref{sec:fusion} and for the retrieval of similar objects, we also interpolate the $g$ and $r$ bands onto a grid of $64$ points over the $100$-day window and normalise each object, so that the grid describes only the shape of the light curve. The $i$ band is too sparse for a grid (Section~\ref{sec:data}) and enters only through the features.

\subsection{Hierarchical classifier}
\label{sec:hier}

We predict in two stages, i.e. a five-way classifier over $\{$SN\,Ia, SN\,CC$^{+}$, AGN, TDE, CV$\}$, where SN\,CC$^{+}$ combines SN\,CC and SLSN, followed by a binary classifier within SN\,CC$^{+}$, so that
\begin{equation}
P(\mathrm{SLSN}) = P(\mathrm{SN\,CC}^{+})\,P(\mathrm{SLSN}\mid\mathrm{SN\,CC}^{+}).
\end{equation}
Both stages are LightGBM models \citep{ke2017lightgbm} with class weights $w_c\propto(1/n_c)^\beta$, and the binary branch is tuned on the area under the curve, because its positive fraction is only $5.7\%$. The hyperparameters are tuned once with Optuna \citep{akiba2019optuna} on a held-out validation split and are then kept fixed. Section~\ref{sec:granularity} evaluates the taxonomy, and Section~\ref{sec:fusion} compares the estimators.

\subsection{External-context dropout}
\label{sec:dropout}

The host photometry, the photometric redshift, the point-source block and the pre-outburst counterpart are unavailable for a fraction of the objects. In order to prevent the model from reading the absence of a block as evidence for a class, each external block is blanked independently on a random $30\%$ of the training rows. This is the photometric counterpart of the metadata dropout of \astranet\ \citep{sasli2026astranet}, which keeps the spectroscopic classifier robust when the metadata of an object are missing, and Section~\ref{sec:missingness} shows why this decision matters.

\subsection{Anomaly layer}
\label{sec:adlayer}

The anomaly layer operates on top of the trained classifier and uses \luna\footnote{\sloppy\luna\ was developed as part of the \astranet\ work, and we refer the reader to \cite{sasli2026astranet} for more details. The code is available at \url{https://github.com/asasli/luna}.}, an out-of-distribution scoring package whose scorers fall into four families \citep[for a taxonomy of such scorers, see][]{yang2021ood,sasli2026astranet}: uncertainty scores from the logits (maximum softmax probability, entropy, energy and ODIN; \citealp{hendrycks2017baseline, liu2020energy, liang2018odin}), distance scores in the embedding space (Mahalanobis variants, $k$-NN and the typicality of the predicted class), density scores (Gaussian mixture, isolation forest \citep{liu2008isolation}, local outlier factor \citep{breunig2000lof} and PCA residual), and a hybrid manifold residual. We use the fourteen scorers that a tree classifier supports, since the two others of \luna\ require a projection head and coarse-level logits. The scorers read an embedding and a set of logits, which are the $160$ features with the raw scores of the tree model, or, for the comparison of Appendix~\ref{app:adgrid} and the retrieval of similar objects, the penultimate layer of a two-branch network over the light-curve grid and the same features.

The anomaly layer reads a narrower feature space than the classifier, i.e. the light-curve features, the host photometry and the Gaia and WISE context, without the photometric-redshift and pre-outburst-counterpart blocks, which help the classification and measurably harm the blind detection (Section~\ref{sec:distance}); the space used is stated next to each result.

The scores are combined either blind, i.e. as the mean percentile rank across the scorers, which requires no labelled anomalies, or by a supervised combiner (logistic regression or LightGBM) when confirmed examples exist. For the blind consensus, the percentile ranks are read from one half of the validation normals and the threshold is set on the other half, so that the score of an object does not depend on the other objects of the batch. The alarm threshold at a nominal rate $\alpha$ is the $\lceil (n+1)(1-\alpha)\rceil$-th order statistic of the $n$ held-out validation normals, a conservative choice that gives an infinite threshold when $n$ is too small, and the false-alarm rate that each threshold realises on the independent test normals is reported next to the main detection rates (Table~\ref{tab:continuum}); it lies between $0.6\%$ and $1.4\%$ for a nominal $1\%$.

\subsection{Evaluation protocols}
\label{sec:protocols}

The classification results come from five-fold cross-validation, reported as the mean and the standard deviation across folds with paired per-fold differences; unless stated otherwise, the quoted $\pm$ values are spreads across folds, draws or orderings, and intervals are given explicitly where a claim needs one. The anomaly detection is evaluated at a false-alarm budget of $1\%$, the operating point that matters when the downstream resource is spectroscopic time, and the human review as recovery against the number of inspected objects. The anomaly experiments use three protocols, which should be compared only within their stated sample: the leave-one-class-out test, which removes a trained class and is optimistic in our experiments (Section~\ref{sec:loco}); the genuine hold-out of $53$ objects from families absent from the taxonomy, the direct but statistically limited test of novelty; and the TDE-free protocol, in which the $57$ TDEs of the pooled training and test splits are hidden targets and the $11$ TDEs of the validation split serve as confirmed examples (Table~\ref{tab:tdecounts} in Appendix~\ref{app:taxonomy}).

\section{Classification}
\label{sec:classification}

\subsection{Classification granularity}
\label{sec:granularity}

We first tuned and evaluated the same pipeline at three levels of granularity. Splitting the core-collapse supernovae into Type II and stripped-envelope supernovae lowers the macro $F_1$ by seven points, and the stripped-envelope objects reach only $F_1 = 0.32$, since they are divided almost evenly between the SN\,Ia and SN\,II classes. Merging the SLSNe into the core-collapse class improves every aggregate metric, because the SLSNe are separated from the Type II supernovae mainly by their absolute magnitude. Both results follow from what a two-band light curve carries, and for this reason the hierarchy of Section~\ref{sec:hier} predicts the resolvable classes at the top level and gives the SLSNe as a ranked branch. A similar hierarchy is used by \astranet\ and by the ORACLE-2 classifiers on the same alert stream \citep{shah2026}.

\begin{table*}[t]
\centering
\caption{Five-fold cross-validated classification as each block of context is added. The upper rows give the accuracy and the balanced accuracy in per cent and the macro $F_1$, with the standard deviation across folds, and the lower rows give the $F_1$ of each class. The accuracy is dominated by the SN\,Ia, which make up $68\%$ of the sample, and for this reason we use the balanced accuracy and the macro $F_1$ as the main metrics.}
\label{tab:classification}
\begin{tabular}{lccccc}
\hline\hline
 & light curve & $+$ host & $+\Mpseudo$ & $+$ point src & $+$ counterpart \\
\hline
accuracy [\%]      & $90\pm1$ & $92\pm1$ & $92\pm1$ & $92\pm1$ & $92\pm1$ \\
balanced acc. [\%] & $71\pm3$ & $79\pm2$ & $80\pm2$ & $81\pm3$ & $83\pm3$ \\
macro $F_1$   & $0.72\pm0.03$ & $0.80\pm0.01$ & $0.80\pm0.01$ & $0.82\pm0.02$ & $0.83\pm0.02$ \\
\hline
SN\,Ia & 0.95 & 0.96 & 0.96 & 0.96 & 0.96 \\
SN\,CC & 0.81 & 0.84 & 0.84 & 0.85 & 0.85 \\
SLSN   & 0.42 & 0.46 & 0.51 & 0.52 & 0.53 \\
AGN    & 0.73 & 0.85 & 0.84 & 0.90 & 0.90 \\
TDE    & 0.58 & 0.75 & 0.75 & 0.76 & 0.81 \\
CV     & 0.84 & 0.93 & 0.93 & 0.93 & 0.94 \\
\hline\hline
\end{tabular}
\end{table*}

\subsection{Host context and the distance chain}
\label{sec:distance}

We add the four context blocks in sequence (Table~\ref{tab:classification}), and we test whether the host location and the distance information reduce the confusions anticipated in Section~\ref{sec:physics} \citep{hammerstein2023tde}. The host photometry gives most of the gain, from a macro $F_1$ of $0.72$ to $0.80$, and the classes that gain are those whose defining property is the position in the host: the TDEs and the AGN, which are nuclear, and the CVs, which have no host. The photometric redshift, through $\Mpseudo$, improves mainly the SLSNe, whose binary branch then ranks them well (area under the curve $0.97$) while the base rate of roughly thirty SLSNe against five hundred core-collapse supernovae per test fold limits the precision at the adopted operating point. The point-source block improves the AGN through the $W1-W2$ colour and the presence of a Gaia counterpart, and the pre-outburst counterpart separates a flare in a nucleus from a supernova near one and marks the CVs that Gaia misses, so that the TDEs and the CVs gain last. The same counterpart block reduces the blind detection of the anomaly layer on the enlarged pool of Appendix~\ref{app:adgrid} from $40\%$ to $35\%$ at a $1\%$ false-alarm budget, and for this reason the anomaly layer reads the features, the host and the Gaia and WISE blocks only.

\begin{figure}[t]
\centering
\includegraphics[width=\columnwidth]{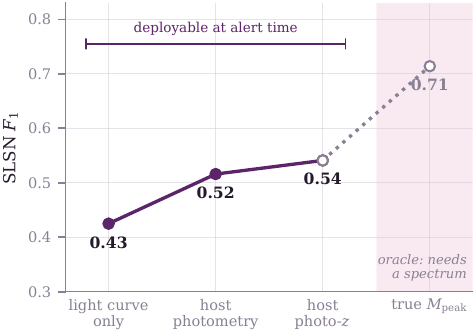}
\caption{Separability of superluminous supernovae against the distance information supplied: five-fold $F_1$ on the four classes with a spectroscopic redshift for every member (the three supernova classes and TDE), under the same hierarchy and folds, so that consecutive points differ only in the distance information. The first three links are available at alert time; the last uses a spectroscopic redshift as a reference for the distance information in this model.}
\label{fig:chain}
\end{figure}

Figure~\ref{fig:chain} isolates the distance information on the four classes with a spectroscopic redshift for every object, which avoids the missingness shortcut of Section~\ref{sec:missingness}: the host photometry and the photometric redshift recover most of the SLSN separation available at the time of the alert, and the spectroscopic distance gives a further gain, which serves as a reference for this model and this feature set (the SLSN $F_1$ values differ slightly from Table~\ref{tab:classification} because the subset differs). The single quantity $\Mpseudo$ performs better than the raw photometric-redshift columns, also on the objects without a redshift.

\label{sec:confusion}

\begin{figure}[t]
\includegraphics[width=\columnwidth]{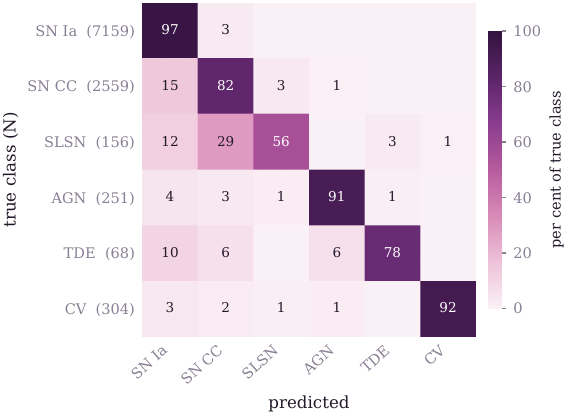}
\caption{Out-of-fold confusion matrix for the headline configuration, in percentages of the true class (sizes in the row labels). The structured leaks are SN\,CC into SN\,Ia, SLSN into SN\,CC, and tidal disruption events divided between the supernova classes and AGN.}
\label{fig:confusion}
\end{figure}

Figure~\ref{fig:confusion} shows the remaining confusions. The largest absolute leak is from the core-collapse supernovae into the SN\,Ia, the precision and the recall of the SLSNe remain close to one half, and the TDE errors enter mainly the supernova classes and the AGN. These residuals are consistent with overlapping light curves and uncertain host or distance information, and the ablations leave open whether other inputs could remove some of them.

\subsection{Calibrated prediction sets}
\label{sec:conformal}

\begin{figure*}[t]
\centering
\includegraphics[width=0.7\textwidth]{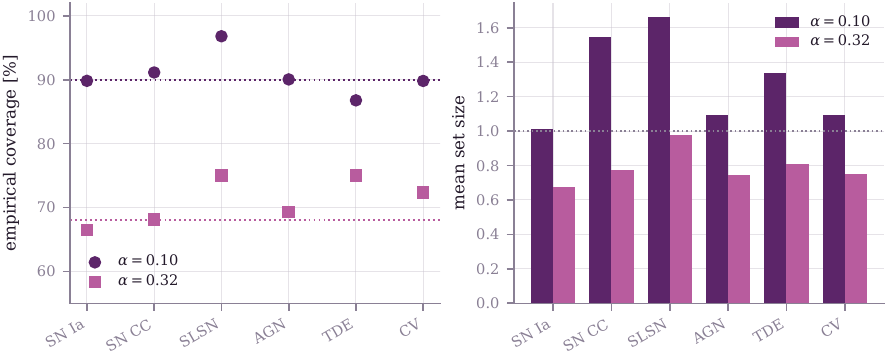}
\caption{Class-conditional conformal prediction on the headline classifier, out of fold. \emph{Left:} empirical per-class coverage against the nominal level for two miscoverage rates. \emph{Right:} mean prediction-set size per true class: an SLSN receives on average $1.7$ candidate labels at $\alpha=0.10$, typically $\{$SLSN, SN\,CC$\}$, mirroring Fig.~\ref{fig:confusion}. Mean sizes below one at $\alpha=0.32$ reflect empty sets.}
\label{fig:conformal}
\end{figure*}

Follow-up schedulers and population studies need calibrated uncertainties in addition to the most probable class. For this reason, the classifier reports class-conditional (Mondrian) conformal prediction sets \citep{vovk2005algorithmic, angelopoulos2021gentle}: for each class, the nonconformity $1 - p_c$ is calibrated separately, and the prediction set contains every class whose probability exceeds its own threshold. The class-conditional construction is important here, since a single marginal guarantee would be dominated by the SN\,Ia and could leave the rare classes under-covered.

At level $\alpha$, a finite class-conditional threshold requires $n_c\ge\lceil1/\alpha\rceil-1$ calibration members, i.e. nine at $\alpha=0.10$, and a simple calibration split leaves only seven TDEs. For this reason, we use the out-of-fold scores of a five-fold cross-fit within each outer training portion; the dropout affects only the fitting rows, and the outer test objects stay outside both fitting and calibration. This construction is related to cross-conformal prediction \citep{vovk2015crossconformal, barber2021jackknife}, and we report the empirical coverage without claiming a finite-sample guarantee. The deployed module uses the same construction over the training and validation splits together.

The result is shown in Fig.~\ref{fig:conformal}. At a nominal coverage of $90\%$, the marginal empirical coverage is $90.2\%$ and $78.9\%$ of the objects receive a single label. The per-class coverages are $89.8\%$, $91.1\%$, $96.8\%$, $90.0\%$, $86.8\%$ and $89.8\%$ for SN\,Ia, SN\,CC, SLSN, AGN, TDE and CV, respectively. Thus the cross-conformal construction does not achieve the nominal level in every class; the TDE estimate is based on only $68$ objects. The classes that the confusion matrix shows to be entangled are exactly the ones that are delivered as candidate sets, which is the appropriate hand-off to spectroscopy.

Each report states the nominal level, the empirical true-class coverage of the predicted class, the credibility, i.e. the largest class $p$-value, and the confidence, i.e. one minus the second-largest $p$-value; the coverage measures how often the set contains the true class, which is a different quantity from the precision of the predicted label. Appendix~\ref{app:diagnostics} shows with one object why the four numbers are reported together.

\subsection{Comparison with sequence models}
\label{sec:fusion}

We compare the trees with the released \applecider\ transformer \citep{applecider2025}, retrained with its own recipe on the same objects, taxonomy and five folds (Appendix~\ref{app:external}), since tabular trees are a strong baseline at this sample size \citep{grinsztajn2022why}. The transformer reaches a macro $F_1$ of $0.64\pm0.03$, against $0.72\pm0.03$ for \flare\ with the light-curve features and $0.84\pm0.02$ with the full context on these folds (the $0.83\pm0.02$ of Table~\ref{tab:classification} comes from the folds of the main comparison). The paired advantage of the light-curve features is $+0.07\pm0.02$, positive in all five folds, with the largest gaps for the AGN and the TDEs, and this comparison supports the adopted features at the present training size.

Averaging the trees with a two-branch convolutional network improves the light-curve-only macro $F_1$ from $0.71$ to $0.76$ (McNemar $p=0.04$), while once the host context and $\Mpseudo$ are included the average no longer improves on the trees ($0.79$ against $0.80$, with $0.76$ for the network alone). This is consistent with an overlap between the information that the network extracts and that of the context, and for this reason we keep the trees alone; the balanced random forest and further tuning are reported in Appendix~\ref{app:external} \citep{sanchezsaez2021alerce, chen2004balanced}.

\section{Anomaly detection}
\label{sec:anomaly}

\subsection{Blind detection of out-of-taxonomy objects}
\label{sec:ceiling}

Without any labelled anomaly, the blind consensus over the fourteen scorers behaves differently on the two halves of the hold-out, and we quote them separately, since an aggregate over the $53$ objects would be dominated by the largest family. The $37$ classical novae reach an area under the curve of $0.92$, and $19\%$ of them (seven objects; $95\%$ interval from $9\%$ to $34\%$) are detected at a false-alarm budget of $1\%$, which realises $1.1\%$ on the test normals. At budgets of $5\%$ and $10\%$, which realise $6.5\%$ and $13\%$, the same consensus detects $57\%$ and $81\%$ of the novae and three and five of the sixteen objects without a parent class, which is the regime in which the review loop of Section~\ref{sec:loop} operates. The sixteen objects from the families without a parent class reach $0.73$, with a single detection, which is too few to quote a rate, and the four families with more than one member lie close to each other, between $0.64$ and $0.75$ (the single FBOT has no area under the curve). The other rates quoted for the novae in this paper come from different protocols, which are listed in Table~\ref{tab:tdecounts}.

This difference agrees with the expectation of Section~\ref{sec:physics}, since the novel families are defined by spectroscopic observables, while the novae have a parent class in the taxonomy (Appendix~\ref{app:taxonomy}). We tested, on identical pools and thresholds, the fourteen scorers alone and in groups, five representations with and without host context, several fusion, routing and cascade rules, and synthetic outlier exposure \citep{hendrycks2019oe}, and the best configurations converge to the same values (Appendix~\ref{app:adgrid}). Three findings explain this result. In the light-curve feature space only the uncertainty scores have probability mass in the extreme tail, so that at a $1\%$ budget the \texttt{energy} score, i.e. the log-sum-exp of the logits, is the only scorer with a usable detection rate. The quantities computed after the softmax lose the total-evidence scale that separates an object matching nothing from an object ambiguous between two classes. The feature-space outlier detectors used for \ztf\ \citep{malanchev2021light} detect one or two objects at this budget. Combining the scorers creates no signal that no single scorer carries, since averaging dilutes the extreme tail, and the CVs, recovered by the density scorers and not by \texttt{energy}, are the one case of complementarity. The representation dominates the blind regime: the host photometry raises the blind detection of the novae from $3\%$ to $19\%$, mainly through the absence of a host. This is what the novae share with the $304$ trained CVs, so that the blind rule detects an object without a host whose light curve does not match the CVs. The limit is unchanged when the anomaly sample is doubled with the $57$ unseen TDEs, when the scorer is chosen with an oracle, and when the $114$ peculiar sub-types are removed from the training set (Appendix~\ref{app:adgrid}). The physical interpretation, i.e. that these families are defined by observables that two-band photometry does not carry, is plausible, but sixteen objects spread over five families are not enough to prove that they cannot be separated.

\subsection{Recovery versus review budget}
\label{sec:loop}

We simulate the active review following \citet{ishida2021active} and \citet{lochner2021astronomaly}, as an operator would carry it out in a vetting platform \citep{vanderwalt2019skyportal, coughlin2023fritz, settles2009active}. The full six-class model with host context ranks the $53$ out-of-taxonomy objects mixed with the $1575$ known-class test objects, and, starting blind, each round inspects the top $20$ remaining candidates, obtains their true labels, refits a logistic combiner on all the inspections and re-ranks the rest. The simulated operator has perfect labels, while a real confirmation may require a spectrum, so that a saving in inspections differs from a saving in telescope time.

\begin{table}[h]
\centering
\footnotesize\setlength{\tabcolsep}{4pt}
\caption{Fraction of the $53$ out-of-taxonomy objects recovered against the number of objects a human inspected, out of a $1628$-object stream. The active loop refits on its own confirmations; the static rankings do not.}
\label{tab:loop}
\begin{tabular}{lccccc}
\hline\hline
review budget & 25 & 50 & 100 & 150 & 250 \\
\hline
static, single scorer  &  9\% & 19\% & 26\% & 32\% & 42\% \\
static, blind consensus & 15\% & 21\% & 34\% & 51\% & 66\% \\
active loop ($K{=}20$) & 15\% & 30\% & \textbf{60\%} & 70\% & \textbf{83\%} \\
\hline\hline
\end{tabular}
\end{table}

\begin{figure*}[t]
\centering
\includegraphics[width=0.7\textwidth]{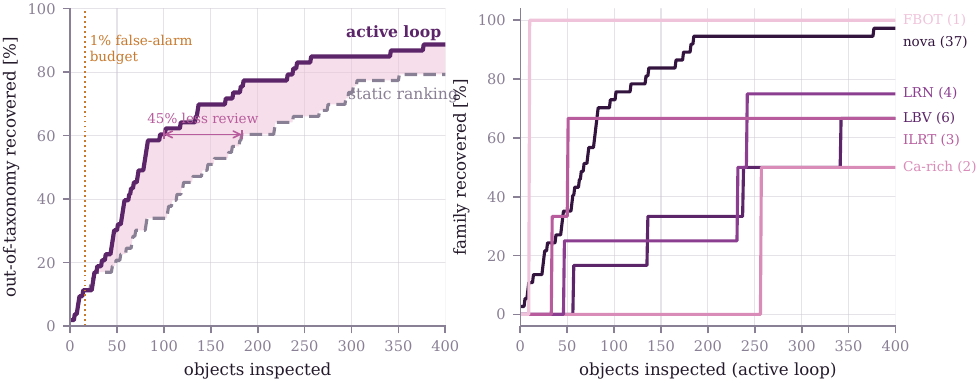}
\caption{Recovery of out-of-taxonomy objects against the number of candidates a human inspected, from a $1628$-object stream containing $53$ of them. \emph{Left:} the active loop against a static blind ranking; the dotted line marks the $1\%$ false-alarm budget. \emph{Right:} the same loop resolved by family.}
\label{fig:loop}
\end{figure*}

Table~\ref{tab:loop} and Fig.~\ref{fig:loop} give the result. Inspecting $6\%$ of the stream recovers $60\%$ of the objects outside the taxonomy, and inspecting $15\%$ recovers $83\%$. The loop reaches with $100$ inspections what the static ranking reaches with $183$, i.e. it saves $45\%$ of the review effort. The novae dominate the early recovery, since with $37$ members the loop learns their signature within a few rounds, while the novel families are recovered more slowly.

The confirmations that the loop collects span several families, and the transfer test shows why this diversity matters. In the TDE-free setup of Section~\ref{sec:lifecycle}, a combiner fitted on eleven confirmed TDEs, i.e. more labels but all from one class, reduces the detection of the real unknowns from $19\%$ to $7\%$, which is consistent with learning the signature of the TDEs, whereas a handful of confirmations from several families improves four of the five held-out families.

\subsection{Retrieval of similar objects}
\label{sec:retrieval}

A confirmed rarity could in principle multiply itself, if the archive were queried for the objects that resemble it and the neighbours were vetted. Retrieval works for the one family with tens of members (for the novae, seven of the twenty nearest neighbours belong to the same family) and returns almost nothing for the families of one to six members. The per-family results are given in Appendix~\ref{app:adgrid}.

\section{From confirmed anomalies to new classes}
\label{sec:lifecycle}

\begin{figure}[t]
\includegraphics[width=\columnwidth]{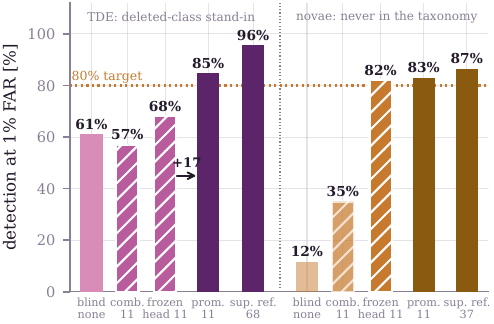}
\caption{The same eleven confirmed members used in three ways, for the TDEs (left) and the classical novae (right). The combiner remains well below the other options in both families, while promotion to a trained class reaches the $80\%$ level (dotted line). The last bar of each family is the supervised class at a fixed false-alarm rate (Table~\ref{tab:continuum}).}
\label{fig:continuum}
\end{figure}

\begin{figure*}[t]
\centering
\includegraphics[width=0.72\textwidth]{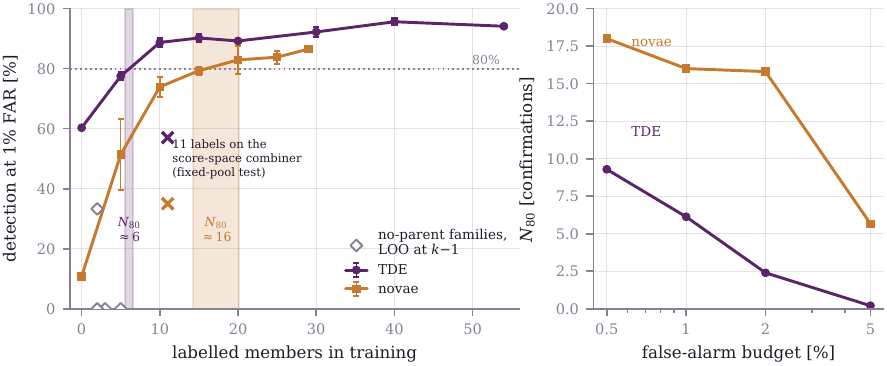}
\caption{Detection at a $1\%$ false-alarm budget against the number of confirmed members in the training set, for the TDEs and the classical novae. Shaded bands mark $N_{80}$, the number of confirmations at which a family reaches $80\%$. Crosses show the same labels spent on the combiner over the anomaly scores, and open diamonds the four families without a parent class. \emph{Right:} $N_{80}$ against the false-alarm budget.}
\label{fig:labelcurve}
\end{figure*}

In this section, we measure what the first confirmations of a new family are worth, depending on where they are used: in a combiner over the anomaly scores, in a head on the frozen features, or as a new class of the classifier (Fig.~\ref{fig:continuum}; the numbers, with the realised false-alarm rates, are in Table~\ref{tab:continuum} of Appendix~\ref{app:replay}). We then vary the number of confirmations for the last of these options (Fig.~\ref{fig:labelcurve}).

For the TDEs (upper block of Table~\ref{tab:continuum} and left panel of Fig.~\ref{fig:continuum}), the combiner over the anomaly scores, which is the standard few-shot approach in the out-of-distribution literature, does not improve on the blind \texttt{energy} rule with eleven labels. However, when the same eleven objects are added to the classifier as a new class, we detect $85\%$ of the hidden TDEs, and five of them already detect $79\%$. With identical labels, targets and budget, the promotion improves on a logistic combiner by $29$ points ($95\%$ interval from $17$ to $42$). A logistic head on the frozen features does better than the combiner, but it still remains well below the promotion.

The TDE-free setup is a deleted-class construction, which flatters the detectors (Section~\ref{sec:loco}), and for this reason we repeat the comparison on the classical novae, which were never part of the taxonomy (lower block of Table~\ref{tab:continuum}). The combiner improves the detection of the novae from $12\%$ to $35\%$, but it remains far below the $82$--$83\%$ of the other two options. A head on the frozen features already matches the promotion, because the properties that define the novae (no host, Galactic position) are directly readable in the feature space. For the TDEs, retraining improves on the logistic head that we tested. For these two families and the models tested, using the confirmations in feature space outperforms combining anomaly scores. Retraining the classifier adds a further gain for the TDEs, while for the novae a head on the frozen features is enough.

The number of confirmations that the promotion needs can also be measured, by thinning each class in the training folds while the test fold keeps all its members (Fig.~\ref{fig:labelcurve}). If we define $N_{80}$ as the number of confirmations at which a family reaches $80\%$ detection, the TDEs need about six and the novae about sixteen, and this ordering holds at every false-alarm budget between $0.5\%$ and $5\%$. These thinning curves use subsamples of the training folds and held-out test folds, while Table~\ref{tab:continuum} uses fixed pools of draws. Therefore, the nova $N_{80}$ and the $83\%$ recovery with eleven labels in that table come from different protocols.

The difference is visible at the left edge of the curves. With no labelled member, the classifier is still trained with one output for the new class, whose probability rises only where every known class scores low (Appendix~\ref{app:adgrid}). This zero-label score already recovers $60\%$ of the TDEs against $11\%$ of the novae. Thus, the TDEs are already isolated by the representation and need only a few labels. On the other hand, the novae are spread across several trained classes (Appendix~\ref{app:taxonomy}) and keep improving well beyond ten. The remaining families are too small for this test, and leave-one-out promotion over the members of the LBVs, LRNe, ILRTs and Ca-rich transients available in the splits ($6$, $4$, $3$ and $3$ objects; Table~\ref{tab:tdecounts} lists which Ca-rich transients enter each experiment) gives one detection in sixteen trials. Blind scoring and a trained class are therefore successive stages of one lifecycle, with the review loop of Section~\ref{sec:loop} moving a family from the first to the second, and the slowest step is the first one, where the first members have to be found one at a time (Section~\ref{sec:retrieval}).

\section{Reliability failure modes}
\label{sec:pitfalls}

In this section, we discuss two evaluation hazards that we encountered, i.e. the class-dependent availability of the context and the use of deleted trained classes as proxies for unknown populations. Each of them would have inflated the reported performance.

\subsection{Class-dependent missingness}
\label{sec:missingness}

Externally sourced features are often unavailable in a class-dependent way. In our catalogue, the spectroscopic redshift is present for every supernova and TDE and for none of the AGN and CVs. This is because of the way the follow-up was allocated and not because of the physics. Supplying this redshift naively to the six-class model raises the macro $F_1$ from $0.74$ to $0.86$, and the AGN $F_1$ from $0.70$ to $0.95$, but essentially all of this gain comes from the model reading a missing value as a class label. Restricted to the four classes where the redshift is always available, the same feature gives a real but much smaller improvement.

The host photometry carries the same hazard in a milder form, since its missingness is physical (a CV has no host galaxy), but the depth of the survey still makes the absence of a detected host unreliable. Without the dropout of Section~\ref{sec:dropout}, the host features improve the objects that have a host and degrade those that do not, with a macro $F_1$ that decreases from $0.78$ to $0.73$ on the latter; with the dropout, they improve both. Reporting the performance separately for the objects with and without each external block is the diagnostic that reveals this effect.

\subsection{Leave-one-class-out tests}
\label{sec:loco}

Deleting a trained class and detecting its members is the standard out-of-distribution benchmark. Choosing a scorer on five deleted classes and applying it to the sixth raises the mean area under the curve from $0.75$ to $0.81$, while the same selection on the out-of-taxonomy families lowers it from $0.62$ to $0.55$, because in this benchmark the deleted classes are coherent and well sampled and the unknown families are small. For this reason, we report both protocols, since the deleted-class performance alone would overstate the discovery performance here.

\section{The observation horizon}
\label{sec:horizon}

\begin{table*}[t]
\centering
\caption{The adopted configuration at truncated horizons, on a fixed split. Left block: classification, with the accuracy in per cent and the $F_1$ values. Right block: the blind anomaly layer on the $53$ out-of-taxonomy objects, given as AUROC / detection [\%] / realised test FPR [\%]. Thresholds target a nominal $1\%$ false-alarm rate using independent calibration objects.}
\label{tab:horizon}
\begin{tabular}{rccccc|cc}
\hline\hline
$h$ [d] & acc. & macro $F_1$ & SLSN & TDE & AGN & consensus & \texttt{energy} \\
\hline
5 & 77 & 0.56 & 0.18 & 0.00 & 0.91 & $0.85/6/0.70$ & $0.63/2/0.57$ \\
10 & 81 & 0.63 & 0.26 & 0.22 & 0.88 & $0.88/23/1.14$ & $0.65/2/0.38$ \\
20 & 86 & 0.73 & 0.48 & 0.43 & 0.90 & $0.90/21/0.44$ & $0.75/0/0.63$ \\
50 & 91 & 0.81 & 0.56 & 0.73 & 0.89 & $0.89/19/0.76$ & $0.83/6/0.63$ \\
100 & 93 & 0.87 & 0.60 & 0.90 & 0.93 & $0.87/13/0.44$ & $0.79/4/0.63$ \\
\hline\hline
\end{tabular}
\end{table*}

\begin{figure*}[t]
\centering
\includegraphics[width=0.8\textwidth]{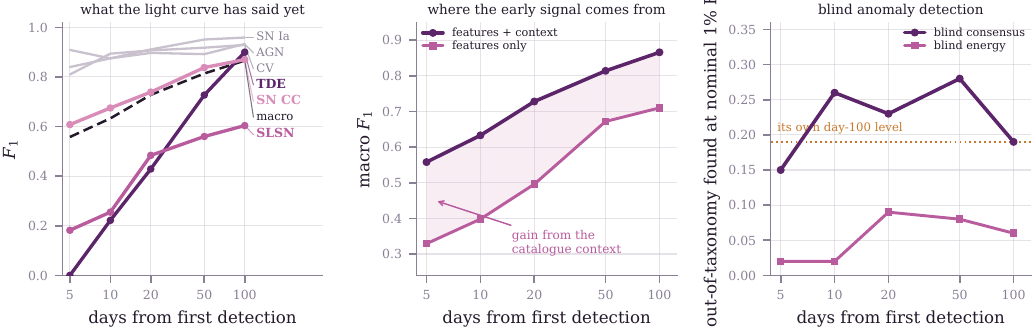}
\caption{Performance against the length of the light curve, for the adopted configuration on a fixed split. \emph{Left:} per-class $F_1$. \emph{Centre:} macro $F_1$ with and without external context; the shaded band is what the context adds. \emph{Right:} blind detection of the out-of-taxonomy objects by the consensus and by the \texttt{energy} rule at a nominal $1\%$ false-alarm threshold. The realised test false-alarm rates appear in Table~\ref{tab:horizon}.}
\label{fig:horizon}
\end{figure*}

All the above results use a window of $100$ days from the first detection, while follow-up decisions are made much earlier. For this reason, we retrain and re-evaluate the adopted configuration on light curves truncated at $5$ to $100$ days, on a fixed split, with every feature and the pseudo-absolute magnitude computed only from the observations inside the window (Fig.~\ref{fig:horizon} and Table~\ref{tab:horizon}). The sample is the same at every horizon, so that the early rows describe objects that later become well-sampled light curves; these results are therefore conditional on the eventual sampling quality, and an unselected early-alert stream remains to be evaluated.

At every horizon the anomaly layer is calibrated as in Section~\ref{sec:adlayer}: $787$ of the $1575$ validation objects define the ranks of the fourteen scores and $788$ set the threshold, with the conservative order statistic of that section, so that the test objects and the rarities enter no stage of it. The deployed module keeps this consensus as its alarm, with $p$-values computed separately per detection-count group and recalibrated on the photometry of the service (Section~\ref{sec:fritz}).

The row at day $100$ differs from Table~\ref{tab:classification} because the horizon study uses one fixed split, on which a single TDE moves the per-class $F_1$ by several points. The AGN are already identified at day $5$ through their catalogue context, while the TDEs and the SLSNe improve as the light curves develop; at day $5$ the context raises the macro $F_1$ from $0.33$ to $0.56$ (Fig.~\ref{fig:horizon}), and the blind consensus reaches its day-$100$ recovery by day $10$ and detects more rarities than the \texttt{energy} rule at every horizon. Since the full-curve model applied to five-day light curves covers the true class for only $55\%$ of the objects at the nominal $90\%$, against $91\%$ for the model trained at that horizon, the deployed module routes each object by the span of its detections to the model trained at $5$, $10$, $20$ or $100$ days, each with its own conformal calibration and anomaly reference population.

\section{Deployment and independent validation}
\label{sec:deployment}

\subsection{Fritz operation}

\begin{figure}[t]
\centering
\includegraphics[width=0.8\columnwidth]{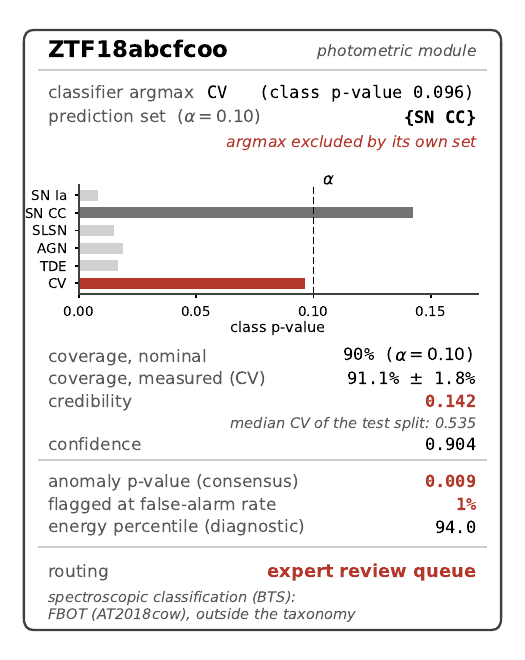}
\caption{The per-object report of the deployed module for the fast blue optical transient ZTF18abcfcoo (AT2018cow), one of the out-of-taxonomy objects of the benchmark, whose $15$ detections over $58$ days route it to the full-curve model. The prediction set at $\alpha = 0.10$ excludes the most probable class, CV (red bar below the $\alpha$ line), the credibility is about a quarter of the median for that class, and the consensus of the anomaly scores flags it at a $1\%$ false-alarm rate ($p = 0.009$), so that the object is sent to expert review. The energy percentile is shown as a diagnostic.}
\label{fig:card}
\end{figure}

The photometric module of \applecider\ \citep{applecider2025} is designed to run inside the \boom\ broker \citep{boom2025}, and it already runs in production on Fritz \citep{vanderwalt2019skyportal, coughlin2023fritz}, the SkyPortal instance of the \ztf\ collaboration, as an analysis service that the platform dispatches to the Open Science Grid for every source saved by the participating programs. For every source, it returns the class probabilities, the prediction set with its four numbers, the anomaly $p$-value of the blind consensus of Section~\ref{sec:anomaly} with its alarms, and a triage verdict (Fig.~\ref{fig:card}). Unlike the single reference used for Table~\ref{tab:horizon}, the $p$-values are computed separately for groups of objects with a similar number of detections, since with a single reference the day-$5$ alarm fired on $10\%$ of the known objects with nine or more detections and on none of those with three or fewer. A classification is posted only when the $90\%$ set contains a single class, or the coarse label supernova when every member of the set is a supernova type. The pipeline takes about $2$\,ms per object on one CPU core in batches and about $35$\,ms per alert on its own, most of it in the fourteen anomaly scorers, against a mean Rubin alert rate of order $300$\,s$^{-1}$. The full model is retrained in minutes, so that the promotion of a new class can follow a nightly cadence.

When the first three ZTF detections of ZTF26abzxgsr, the optical counterpart of a gamma-ray burst, which were taken within an hour and a half, are replayed through the current version of the module, the consensus flags the object at a $5\%$ per-analysis false-alarm rate ($p = 0.015$), and its lifetime statistic reaches the monthly threshold of Section~\ref{sec:far}, at which $12$ of the $1\,882$ known objects are as significant, i.e. one false alarm per $21$ days on the stream, against one per $14$ days with the benchmark calibration. Its prediction set, \{SN\,CC, CV\}, contains two classes, so that no classification is posted. This is a retrospective replay on the photometry that Fritz holds, and since this object also entered the choice of Section~\ref{sec:fritz} to keep the boosters, it illustrates the alarm and does not confirm it independently.

\subsection{Transfer to the photometry of the service}
\label{sec:fritz}

The ZTF photometry supplied by Fritz is denser than the public-alert
light curves used to build the benchmark through \boom. For the
benchmark test objects, it contains a median of $93$ detections within
$100$ days, compared with $26$ in the benchmark; only $39\%$ of its
$g$- and $r$-band epochs are shared with the benchmark, and $i$-band
detections are available for $45\%$ of the objects.
This difference in the input affects both classification and anomaly
calibration. On the same $1\,555$ BTS test objects, routed as in the
service, the macro $F_1$ falls from $0.836$ to $0.807$, while the
realised false-alarm rate at the nominal $1\%$ rises from $0.6\%$
to $2.4\%$.

We evaluated two responses to this shift: retraining the models on
the Fritz photometry, and retaining their weights while recalibrating the anomaly layer. Full retraining restored the false-alarm rates without a clear improvement in classification across seeds, but no longer flagged the gamma-ray burst discussed in Section~\ref{sec:deployment}. We therefore retained the classifiers and anomaly boosters and recalibrated the anomaly layer on the photometry supplied by Fritz. For each of the four phase models, the consensus ranks and detection-count-dependent $p$-values use separate halves of the validation split. The test objects enter neither stage. On the $1\,882$ BTS and Fritz test objects, the recalibrated alarm realises false-alarm rates of $0.6\%$, $3.3\%$ and $8.2\%$ at nominal levels of $1\%$, $5\%$ and $10\%$, respectively, with unchanged classifications.

This recalibration concerns the anomaly alarm, not the classification prediction sets. The full-curve classifier's nominal $90\%$ sets contain the true class for only $83\%$ of the Fritz test objects. The benchmark coverage reported in Section~\ref{sec:conformal} therefore does not transfer directly to the service population. The prediction sets require their own calibration on representative Fritz sources.

\subsection{False-alarm rate per unit time}
\label{sec:far}

Repeated analyses increase the number of false alarms that reach
the reviewer. In our replay of the Fritz photometry, a nominal
$1\%$ per-analysis alarm flags about $10\%$ of the known-class
sources at least once within $100$ days. At the observed arrival
rate of $6.8$ saved sources per day, this corresponds to roughly
five false alarms per week. A threshold for a single analysis
therefore does not describe the workload of a service that
revisits each source as its light curve grows.

To express that workload directly, we calibrate the alarm over
the full sequence of analyses and report its false-alarm rate
per unit time, following the background-rate interpretation
used in gravitational-wave searches \citep{abbott2016gw150914}.
The lifetime statistic $P_{\mathrm{life}}$ summarises the most
significant departure from the known population over the
observed history of a source. Its calibration uses replayed
known-class light curves: the fraction whose whole-life
statistic is at least as extreme, multiplied by the source
arrival rate, gives the estimated false-alarm rate. This is an
empirical calibration for the replay's observing schedule and
$100$-day horizon, not an anytime-valid $p$-value
\citep{johari2022alwaysvalid}.
Appendix~\ref{app:replay} gives the construction and evaluation
protocol.

We evaluated thresholds targeting one false alarm per week,
per month and per quarter using two-fold cross-fitting.
Each known-class object was evaluated with a reference and
thresholds constructed from the opposite half of the sample.
The three thresholds flagged $39$, $14$ and $11$ of the
$1\,882$ objects, respectively. At the measured arrival rate,
these counts correspond to approximately one false alarm every
$7$, $20$ and $25$ days. The weekly threshold therefore matches
its intended rate in this replay. The monthly estimate is
consistent with its target within the sampling uncertainty,
but the quarterly threshold does not achieve its target
(Appendix~\ref{app:replay}). These time intervals also depend
on the source arrival rate and must be updated if the stream
changes.

\begin{figure*}[t]
\centering
\includegraphics[width=0.85\textwidth]{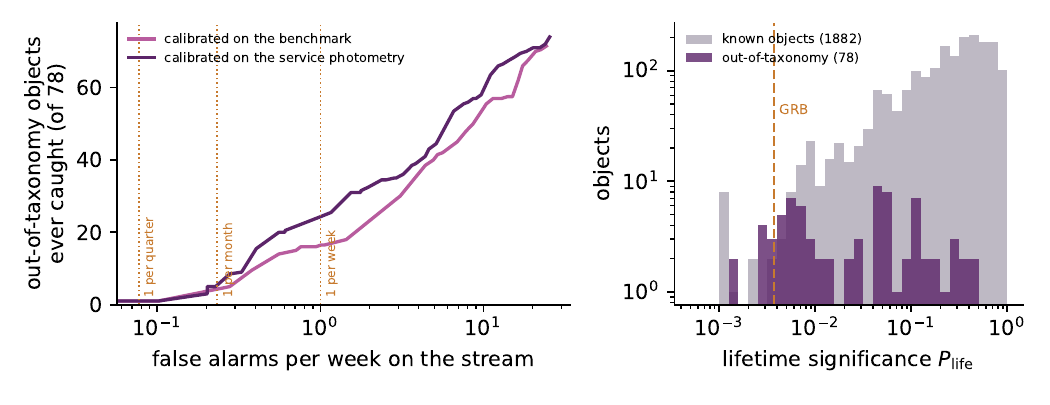}
\caption{The alarm of the deployed module as a false-alarm rate per unit time, on the $1\,882$ known-class test objects and the $78$ out-of-taxonomy objects with the photometry that the service receives, replayed detection by detection. \emph{Left:} out-of-taxonomy objects caught at some point in their life against the false alarms per week that the same threshold on $P_{\mathrm{life}}$ produces on the known objects (reference from one half of the known objects, rates on the other), for the alarm calibrated on the benchmark light curves and on the photometry of the service; the dotted lines mark one false alarm per week, per month and per quarter at the measured rate of $6.8$ sources per day. \emph{Right:} the distribution of $P_{\mathrm{life}}$ for the known and the out-of-taxonomy objects with the second calibration, and the value of the gamma-ray burst ZTF26abzxgsr at its first analysis.}
\label{fig:far}
\end{figure*}

The service uses the weekly threshold to send sources for
expert review. With the shipped reference, this threshold
recovers $27$ of the $78$ out-of-taxonomy objects
($35\%$, with a $95\%$ interval of $25$--$46\%$), predominantly
novae and two fast blue optical transients. The nominal
monthly and quarterly thresholds recover $13$ and $4$ objects.
The spectroscopically defined families remain undetected,
consistent with the benchmark results (Fig.~\ref{fig:far}).

\subsection{Transfer of the learning from confirmations}
\label{sec:transferlabels}

The two supervised steps of Sections~\ref{sec:loop} and~\ref{sec:lifecycle} transfer to this stream as well. The review loop, run on the same $1\,960$ objects with the Fritz photometry and the same protocol, recovers $42\%$ of the $78$ out-of-taxonomy objects after $100$ inspections and $71\%$ after $250$, against $37\%$ and $55\%$ for the static ranking by the consensus. Therefore, the gain of the loop appears later than on the benchmark and reaches the same size. A logistic combiner over the fourteen scores, fitted on the known objects of the validation split and on the $50$ out-of-taxonomy objects of the benchmark and applied to the $28$ Fritz objects that it never saw, detects $17$ of them at a realised $5\%$ against $11$ for the blind consensus, with the same three detections outside the novae. This transfer of the first confirmations between populations is what the deployed module uses to order its review queue.

\section{Discussion}
\label{sec:discussion}

In the alert stream, the photometric ranking complements the image-based filters that select the real and bright transients \citep{duev2019braai, rehemtulla2024btsbot}. In our benchmark, about one in three of the first hundred candidates of the consensus ranking lies outside the taxonomy, against one in thirty-one in the whole evaluation stream. Active review further raises this yield, and spectroscopy confirms the candidates. A promoted class also needs at least nine calibration members before its class-conditional threshold at $\alpha=0.10$ becomes finite, which is distinct from the number of labels that its detection needs.

The comparison of the uses of the labels is the central practical result of this work. Both families benefit more from learning in the feature space than from combining the anomaly scores, and only the TDEs show a clear additional gain from retraining over the frozen-feature head that we tested. The recovery-versus-budget curves and $N_{80}$ are useful measures for repeating this comparison in other surveys, while their numerical values describe the present samples and protocols.

The main limitations come from the size and the selection of the samples. The rarest families contain one to six objects, too few to establish a general photometric detection limit, and the review loop assumes perfect labels at a prevalence of $3.3\%$, much higher than in a raw alert stream, where the first rounds would return fewer confirmations. The early-time study selects the objects by their eventual light-curve quality, and the transfer to the photometry of the service shows that both the false-alarm rates and the prediction-set coverage change with the input (Sections~\ref{sec:fritz} and~\ref{sec:far}). A larger hold-out, for example from the Transient Name Server \citep{tns}, would require a matched reference population, and the six-band light curves of the Rubin Observatory \citep{ivezic2019lsst}, as well as sequence models pre-trained on much larger unlabelled samples, may change the comparisons of this work.

Finally, the first population on which the method can produce new results is the $7\,631$ BTS objects that never received a spectrum, which are half a magnitude fainter than the benchmark and were selected by the same survey under the same criteria. The objects whose prediction set excludes their own most probable class and the top of the anomaly ranking form a target list for spectroscopy, once the false-alarm threshold has been recalibrated on classified objects of matched magnitude, and we leave these results to a companion paper.

\section{Conclusions}
\label{sec:conclusions}

We have introduced \flare, which combines photometric classification, anomaly ranking and learning from confirmed candidates, and we have followed transient novelty from blind detection through human review to supervised classification on one stream of \ztf\ data. Catalogue context raises the classification where the defining property of a class is absent from the light curve; blind detection recovers few objects of the smallest unknown families, while a simulated review loop recovers most of the hold-out for a small fraction of the inspections; and the first confirmations of a new family are most useful in the feature space, with an additional gain from retraining that depends on the family.

These results suggest a practical sequence: rank the unlabelled candidates for review, and learn from their features as the confirmations accumulate. The deployment on Fritz shows that the module operates in production, and the transfer to the photometry of the service shows that the calibration has to be made on that photometry and that an alarm of a service which re-analyses its sources is best reported as a false-alarm rate per unit time. We expect that this framework will be very useful for the Rubin era, where the path of an unknown object from the first alarm to a trained class has to be short.

\section{Data and code availability}
The benchmark, i.e. the light curves, the table of $160$ features, the host photometry, the photometric redshifts, the Gaia and WISE point-source context, the Pan-STARRS pre-outburst counterpart block, the splits and the out-of-taxonomy hold-out, together with the classifier and the analysis code, are available at \url{https://github.com/applecider-ml/flare}. The out-of-distribution scoring package used by the anomaly layer is developed separately and is available at \url{https://github.com/asasli/luna}.

\section*{Acknowledgments}
We thank David Cook for his review of the manuscript on behalf of the ZTF Publications Board and for useful comments.

The UMN authors acknowledge support from the National Science Foundation with grant numbers PHY-2117997, PHY-2308862 and PHY-2409481. A.S. and M.C. acknowledge support from LSST-DA through grants 2025-SFF-LFI-14-Coughlin, 2026-SFF-LFI-17-Sasli, and 2026-LSSTDA-SFF-Events-11-Sasli. LINCC Frameworks is supported by Schmidt Sciences, a philanthropic initiative founded by Eric and Wendy Schmidt, as part of the Virtual Institute of Astrophysics (VIA).

Based on observations obtained with the Samuel Oschin Telescope 48-inch and the 60-inch Telescope at the Palomar Observatory as part of the Zwicky Transient Facility project. ZTF is supported by the National Science Foundation under Grants No. AST-1440341, AST-2034437, and currently Award \#2407588. ZTF receives additional funding from the ZTF partnership. Current members include Caltech, USA; Caltech/IPAC, USA; University of Maryland, USA; University of California, Berkeley, USA; Cornell University, USA; Drexel University, USA; University of North Carolina at Chapel Hill, USA; Institute of Science and Technology, Austria; National Central University, Taiwan, German Center for Astrophysics, Germany, and OKC, University of Stockholm, Sweden. Operations are conducted by Caltech's Optical Observatory (COO), Caltech/IPAC, and the University of Washington at Seattle, USA.

The Pan-STARRS1 Surveys (PS1) and the PS1 public science archive have been made possible through contributions by the Institute for Astronomy, the University of Hawaii, the Pan-STARRS Project Office, the Max-Planck Society and its participating institutes, the Max Planck Institute for Astronomy, Heidelberg and the Max Planck Institute for Extraterrestrial Physics, Garching, The Johns Hopkins University, Durham University, the University of Edinburgh, the Queen's University Belfast, the Harvard-Smithsonian Center for Astrophysics, the Las Cumbres Observatory Global Telescope Network Incorporated, the National Central University of Taiwan, the Space Telescope Science Institute, the National Aeronautics and Space Administration under Grant No.\ NNX08AR22G issued through the Planetary Science Division of the NASA Science Mission Directorate, the National Science Foundation Grant No.\ AST-1238877, the University of Maryland, E\"otv\"os Lor\'and University (ELTE), the Los Alamos National Laboratory, and the Gordon and Betty Moore Foundation.

The Legacy Surveys consist of three individual and complementary projects: the Dark Energy Camera Legacy Survey (DECaLS; NOIRLab Proposal ID \#2014B-0404; PIs: David Schlegel and Arjun Dey), the Beijing-Arizona Sky Survey (BASS; NOIRLab Proposal ID \#2015A-0801; PIs: Zhou Xu and Xiaohui Fan), and the Mayall $z$-band Legacy Survey (MzLS; Proposal ID \#2016A-0453; PI: Arjun Dey). DECaLS, BASS and MzLS together include data obtained, respectively, at the Blanco telescope, Cerro Tololo Inter-American Observatory, NSF's NOIRLab; the Bok telescope, Steward Observatory, University of Arizona; and the Mayall telescope, Kitt Peak National Observatory, NOIRLab. The Legacy Surveys project is honored to be permitted to conduct astronomical research on Iolkam Du'ag (Kitt Peak), a mountain with particular significance to the Tohono O'odham Nation.

This work has made use of data from the European Space Agency (ESA) mission Gaia (\url{https://www.cosmos.esa.int/gaia}), processed by the Gaia Data Processing and Analysis Consortium (DPAC, \url{https://www.cosmos.esa.int/web/gaia/dpac/consortium}). Funding for the DPAC has been provided by national institutions, in particular the institutions participating in the Gaia Multilateral Agreement.

This publication makes use of data products from the Wide-field Infrared Survey Explorer, which is a joint project of the University of California, Los Angeles, and the Jet Propulsion Laboratory/California Institute of Technology, and NEOWISE, which is a project of the Jet Propulsion Laboratory/California Institute of Technology. WISE and NEOWISE are funded by the National Aeronautics and Space Administration. This research has made use of the NASA/IPAC Infrared Science Archive, which is funded by NASA and operated by the California Institute of Technology.

This research made use of \textsc{Astropy}, LightGBM, Optuna, and the \texttt{light-curve} feature-extraction library.

\textbf{\textit{AI disclosure statement.}} During the preparation of this manuscript, the authors used Claude (Fable 5.1 and Opus 5.5) and ChatGPT (OpenAI) to assist with editing and restructuring portions of the text for clarity. The authors reviewed and revised the AI-assisted text and take full responsibility for the content of the manuscript.

\appendix

\section{Taxonomy}
\label{app:taxonomy}

Table~\ref{tab:taxonomy} gives the mapping from the spectroscopic sub-types to the trained classes. The grouping is a coarsening of the \astranet\ taxonomy \citep{sasli2026astranet}, with its Ibc and II classes merged into SN\,CC, so that the spectroscopic probabilities map onto these classes by summation. The peculiar members of a trained class (SN\,Iax, Ibn, Ia-CSM, SLSN-II and so on) are mapped to their parent class, since $176$ such objects are already present in the labelled sample under their parent type, and scoring them as novelties would penalise correct classifications.

\begin{table}[h]
\centering
\footnotesize\setlength{\tabcolsep}{4pt}
\caption{Sub-type to class mapping.}
\label{tab:taxonomy}
\begin{tabular}{ll}
\hline\hline
class & spectroscopic sub-types \\
\hline
SN\,Ia & Ia, Ia-91T, Ia-91bg, Ia-pec, Iax, Ia-CSM, Ia-SC \\
SN\,CC & II, IIP, IIL, IIn, IIb, II-pec, Ib, Ic, Ic-BL, Ibn, Icn \\
SLSN   & SLSN-I, SLSN-II \\
AGN    & AGN \\
TDE    & TDE, TDE-H-He, TDE-He, TDE-featureless \\
CV     & CV \\
\hline
\multicolumn{2}{l}{\emph{held out, no parent class}} \\
novel  & LBV, LRN, ILRT, Ca-rich, FBOT \\
novae  & classical nova \\
\hline\hline
\end{tabular}
\end{table}

The classical novae are reported separately from the novel families. A nova is a thermonuclear outburst on the white dwarf of a CV system, so that it has a parent class, but photometrically the two are different populations: only $32\%$ of the $37$ held-out novae are classified as CVs, with a median $P(\mathrm{CV})$ of $0.14$ against $0.99$ for the CVs of the test split, and a median credibility less than half that of a CV, at similar brightness and duration. Merging them would add a distinct population to the cleanest class of the confusion matrix, while treating them as a family without a parent class would mix two detection problems.

\begin{table}[t]
\centering
\footnotesize\setlength{\tabcolsep}{4pt}
\caption{The subsets of the TDEs and the classical novae that each experiment uses. The nova rows also give the corresponding rate, since the blind numbers differ in both method and sample.}
\label{tab:tdecounts}
\begin{tabular}{lr}
\hline\hline
experiment & objects \\
\hline
\multicolumn{2}{l}{\emph{TDEs: $68$ (train $47$ / val $11$ / test $10$)}} \\
five-fold classification, out of fold & all $68$ \\
thinning curve, per training fold (at most) & $54$ \\
TDE-free setup: hidden targets (train $+$ test)     & $57$ \\
few-shot arms: confirmed members (validation)       & $11$ \\
promotion arms (drawn from the validation $11$)     & $5$ or $11$ \\
leave-one-class-out test fold                       & $10$ \\
\hline
\multicolumn{2}{l}{\emph{Ca-rich transients}} \\
out-of-taxonomy hold-out (BTS type Ca-rich)          & $2$ \\
leave-one-out promotion (BTS type SN Ca-rich-Ca, & \\
\quad filed under SN\,CC, removed from the normals)  & $3$ \\
\hline
\multicolumn{2}{l}{\emph{classical novae: all $37$ in the anomaly hold-out}} \\
zero-label score (Fig.~\ref{fig:labelcurve}): $11\%$ & all $37$ \\
blind \texttt{energy}, Table~\ref{tab:continuum}: $12\pm3\%$   & $26$ targets \\
blind consensus (Sec.~\ref{sec:ceiling}): $19\%$ & all $37$ \\
promotion, out of fold, full complement: $87\%$     & all $37$ \\
\hline\hline
\end{tabular}
\end{table}

\section{Scorer grid, robustness controls and retrieval}
\label{app:adgrid}

This appendix contains the material behind Sections~\ref{sec:ceiling} and~\ref{sec:retrieval}.

\begin{figure}[t]
\centering
\includegraphics[width=0.5\columnwidth]{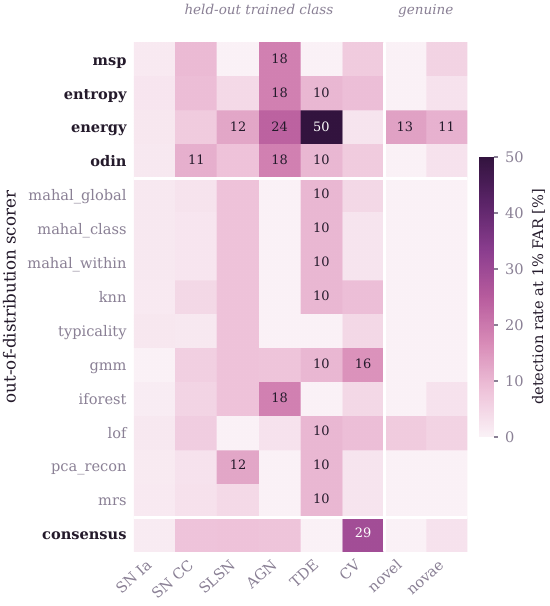}
\caption{Detection at a $1\%$ false-alarm budget for every scorer and every hidden class, in the light-curve feature space before the host block is added. Columns to the left of the divider are leave-one-class-out tests; the two on the right are the out-of-taxonomy hold-out. On the out-of-taxonomy hold-out, only the uncertainty scores (top four rows) have appreciable mass in the extreme tail, and \texttt{energy} carries most of it, with the CVs as the one exception, since they are found through the density scores and the consensus.}
\label{fig:heatmap}
\end{figure}

Figure~\ref{fig:heatmap} shows the detection of every scorer for every hidden class. Averaging the scorers dilutes the extreme tail, so that the mean fusion detects nothing at the $1\%$ budget although its area under the curve is high. The best cascade only ties the single best scorer, and the routing by predicted class is worse than a global rule, because the subsets on which it conditions are too small. The CVs show what a real complementarity looks like: they are recovered by the consensus at $29\%$ and by the \texttt{energy} score at $2\%$, since they are confidently classified but occupy an unusual region of the feature space.

Two controls test whether the limit of Section~\ref{sec:ceiling} is an artefact of our setup. Retraining without the $114$ peculiar sub-types filed under their broad classes leaves both the detection of the $25$ peculiar test members and the out-of-taxonomy results unchanged, since these sub-types resemble their parent class in photometry. Adding the $57$ unseen TDEs to the $53$ rarities doubles the anomaly sample, on which the best blind rule detects $40\%$, with $61\%$ of the TDEs, and a subset of scorers chosen on the targets themselves does no better than the \texttt{energy} rule fixed in advance.

\begin{table}[h]
\centering
\footnotesize\setlength{\tabcolsep}{4pt}
\caption{Same-family retrieval in the embedding of the network. The last column gives, for one confirmed member, how many of its siblings appear among its twenty nearest neighbours, averaged over the choice of member.}
\label{tab:retrieval}
\begin{tabular}{lccc}
\hline\hline
family & $N$ & $p@10$ & siblings in top 20 \\
\hline
novae   & 37 & $0.43$ & $7.4$ of $36$ \\
ILRT    &  3 & $0.07$ & $1.0$ of $2$ \\
Ca-rich &  2 & $0.00$ & $0$ \\
LBV     &  6 & $0.00$ & $0$ \\
LRN     &  4 & $0.00$ & $0$ \\
FBOT    &  1 & --- & --- \\
\hline\hline
\end{tabular}
\end{table}

The zero-label score of Section~\ref{sec:lifecycle} follows from the training of the trees. With one output for the new class and no member of it, the probability of that class is
\begin{equation}
    P_{\rm new} = \frac{e^{z_{\rm new}}}{\sum_k e^{z_k}},
\end{equation}
where $z_k$ are the raw scores of the trees. The raw score $z_{\rm new}$ starts from the boosting initialisation and is penalised on the known-class training objects since its gradient under the cross-entropy loss is $P_{\rm new}$ itself for every training object. Therefore, it stays low wherever the training data lie and $P_{\rm new}$ rises only where all the known classes also score low.

For the retrieval of Section~\ref{sec:retrieval}, measured with cosine nearest neighbours in the embedding of the network over the same stream, the precision at ten neighbours for same-family retrieval is $0.31$ against a chance rate of $0.016$, but Table~\ref{tab:retrieval} shows that this comes from one family. A confirmed nova returns seven more novae among its twenty nearest neighbours, while a confirmed LBV returns none, so that the families that most need to grow are those too sparse for their members to be each other's neighbours.

\section{Model comparisons}
\label{app:external}

This appendix contains the comparisons behind Section~\ref{sec:fusion}.

The \applecider\ photometric classifier was retrained with the configuration of its release \citep{applecider2025}. Each detection becomes a token built from the time since the first detection and since the previous detection (both as $\log(1+\Delta t)$), the log-flux and its uncertainty, and a one-hot band. The times also enter through a Time2Vec embedding, and the encoder has four layers of dimension $128$ with eight heads, a class token and a dropout of $0.3$. On each fold, the encoder is first pre-trained for $50$ epochs to reconstruct $30\%$ of masked detections, and is then fine-tuned with a focal loss ($\gamma=2$), a class-balanced sampler and early stopping on the same early-stopping objects that the trees use. The only changes from the release are the six classes of this work in place of five and the $100$-day window. The per-class $F_1$ is $0.93$ for the SN\,Ia, $0.74$ for the SN\,CC, $0.36$ for the SLSNe, $0.59$ for the AGN, $0.45$ for the TDEs and $0.81$ for the CVs.

Copying a number from the published \alerce\ light-curve classifier \citep{sanchezsaez2021alerce, forster2021alerce} would confound the taxonomy, the sample selection and the photometry with the estimator, and for this reason we replace only the estimator, i.e. we train the balanced random forest \citep{chen2004balanced, breiman2001random} that \alerce\ uses, in a flat and in a balanced-hierarchical arrangement, on the identical folds, features, context blocks and dropout. Both forests reach a macro $F_1$ of $0.74\pm0.02$, against $0.83\pm0.02$ for the boosted trees, and the difference is concentrated on the rare classes (a TDE $F_1$ of about $0.65$ against $0.81$, and an SLSN $F_1$ of about $0.32$ against $0.53$). With the same hyperparameter search that our trees received, the hierarchical forest improves to $0.76\pm0.02$, which still leaves a deficit that is positive in every fold.

Finally, the trees themselves leave little room for improvement: a five-seed ensemble and a new hyperparameter search change the macro $F_1$ by at most $0.002$, adding nine Gaussian-process summaries of the kind used in the PLAsTiCC challenge \citep{boone2019avocado, kessler2019plasticc, hlozek2023plasticc} lowers it slightly, and the fusion with a dilated-convolution (TCN) branch becomes slightly positive only once the point-source block is included ($+0.008\pm0.011$), well within the fold-to-fold spread.

\section{Catalogue construction}
\label{app:context}

For the host photometry, we query Pan-STARRS DR2 \citep{chambers2016ps1} within $30''$ of every position and record the nearest extended source ($i_{\mathrm{PSF}} - i_{\mathrm{Kron}} > 0.05$), i.e. its angular offset, its Kron magnitudes and colours, and the number of extended neighbours. Of the $10\,497$ positions, $10\,487$ return a field and $10\,229$ contain an extended source. The resulting statistics follow the physics of the classes: TDEs and AGN lie on their host (median offsets of $0.12''$ and $0.14''$), supernovae are displaced by $3$--$4''$, the hosts of SLSNe are two magnitudes fainter than the average host, and only $77\%$ of the CVs have any extended source nearby, at a median separation of $15''$. This nearest-source rule will assign the wrong host to some offset transients near bright galaxies; probabilistic associations, such as the host matching that the Transient Name Server now publishes \citep{yaron2026hosts}, are better, but they cannot be substituted into these columns, since the Legacy Surveys model magnitudes are $0.25$\,mag brighter than the Pan-STARRS Kron magnitudes of the same galaxy (scatter $0.19$\,mag) and the two catalogues select a different nearest galaxy for a quarter of the sample, so that adopting them requires refitting the host block.

Photometric redshifts come from Legacy Surveys DR10, with DR9 as the northern fallback \citep{dey2019legacy, zhou2021photoz}. We select the nearest extended source carrying a photo-$z$: $8891$ positions return a field and $7582$ yield a usable value. Equation~\eqref{eq:mpseudo} converts the observed peak to $\Mpseudo$. The median photo-$z$ is $0.32$ for SLSNe and $0.05$ for core-collapse supernovae, consistent with the larger accessible volume for luminous events.

For the point-source context, Gaia DR3 \citep{gaia2023dr3} within $2''$ provides the parallax and the proper motion with their significances, the $G$ magnitude, the RUWE, and, more informatively than any of these quantities, the presence of a point source at all: $98\%$ of the AGN and $84\%$ of the TDEs have a Gaia counterpart (a nuclear point source), against $8$--$21\%$ of the supernovae and $18\%$ of the CVs, whose quiescent counterparts mostly lie below the Gaia limit. AllWISE \citep{wright2010wise} within $6''$ provides the mid-infrared colour $W1-W2$, which is the classical AGN discriminant \citep{stern2012wise}: $42\%$ of the AGN have $W1-W2>0.8$ against at most $2\%$ of every other class.

For the pre-outburst counterpart, Pan-STARRS reaches about $2.5$ magnitudes deeper than the point-source limit of Gaia and its $3\pi$ survey of 2010--2014 \citep{chambers2016ps1} precedes every transient in the sample, so that a source within $1.5''$ of the position is a candidate counterpart (the quiescent binary of a CV, a massive progenitor or a persistent nucleus); the same DR2 query provides it, and $41\%$ of the positions carry one. All four blocks are externally sourced and unevenly available, which makes them the natural setting for the missingness failure mode of Section~\ref{sec:missingness}.

\section{Replay protocol and the use of the first labels}
\label{app:replay}

The replay of Section~\ref{sec:far} analyses every light curve at every detection up to the fortieth and at every fifth detection afterwards; each prefix is routed to the phase model that its span selects, and its per-analysis $p$-value $p_k$ comes from the consensus of that model, calibrated on the validation split (Section~\ref{sec:fritz}). At the $k$-th analysis, the running minimum is $m_k = \min_{j\le k} p_j$, and its lifetime $p$-value is
\begin{equation}
\pi_k = \frac{1 + \#\{i : m^{\mathrm{ref}}_{i,k} \le m_k\}}{1 + N_k},
\end{equation}
where $m^{\mathrm{ref}}_{i,k}$ are the running minima of the $N_k$ known reference objects that reach the $k$-th analysis (beyond the analysis index that at least $50$ of them reach, the last such index is used); the lifetime statistic is $P_{\mathrm{life}} = \min_k \pi_k$ over the analyses so far. The false-alarm rate reported for an object is $(1 + \#\{i : P^{\mathrm{ref}}_{\mathrm{life},i} \le P_{\mathrm{life}}\})/(1+N)$, computed over the whole-life statistics of the $N$ known reference objects, times the rate at which sources enter the service, and the threshold of each level is the corresponding quantile of $P^{\mathrm{ref}}_{\mathrm{life}}$. The schedule depends only on the number of detections, so that a new detection adds analyses without removing any and an alarm reached at an earlier analysis is kept; beyond the fortieth detection, the newest detections enter $P_{\mathrm{life}}$ at the next multiple of five. At analysis time the deployed module replays the prefixes of the light curve with the same schedule, which takes about two seconds, so that the service keeps no state between analyses. The reference that the module ships uses all $1\,882$ known test objects, so we verified the levels of Section~\ref{sec:far} by cross-fitting on two halves: the running-minimum reference and the thresholds are built from one half of the test objects, each object of the other half is evaluated once, under a reference that excludes it, and the two held-out halves together give one evaluation of every test object. Since the per-analysis $p$-values come from the validation split alone, no stage of the calibration sees the evaluated object. The pooled held-out counts of Section~\ref{sec:far}, $39$, $14$ and $11$ of $1\,882$, have approximate Wilson intervals of $1.5$--$2.8\%$, $0.4$--$1.2\%$ and $0.3$--$1.0\%$, approximate because the two folds share the objects in their reference and evaluation roles, so that the quarterly interval excludes its target of $0.2\%$.

Finally, Table~\ref{tab:continuum} gives the numbers behind Fig.~\ref{fig:continuum}, with the false-alarm rate that each threshold realises on the independent test normals.

\begin{table}[h]
\centering
\footnotesize\setlength{\tabcolsep}{4pt}
\caption{Detection at a $1\%$ false-alarm budget ($\tpr$) when the same confirmed members are used in different ways, and the false-alarm rate that the threshold realises on the independent test normals (FPR). Upper block: the $57$ TDEs that the model never saw, hidden among $1565$ test objects. Lower block: the $37$ classical novae, with $11$ confirmed and $26$ targets over twenty draws. The last row of each block ($\dagger$) is the supervised class, measured out of fold over all members, with the threshold set on the out-of-fold scores of the other objects, so that its false-alarm rate is $1\%$ by construction. These two rows are therefore a reference performance at an empirically fixed false-alarm rate, whereas every other row uses a threshold set on the validation normals and evaluated on the independent test normals. Spreads are over draws or orderings of the labels.}
\label{tab:continuum}
\begin{tabular}{rlcc}
\hline\hline
labels & use of the labels & $\tpr$ & FPR [\%] \\
\hline
\multicolumn{4}{l}{\emph{TDEs}} \\
none & blind \texttt{energy} & $0.61$ & $1.3$ \\
$11$ & combiner over scorers & $0.57$ & $1.4$ \\
$11$ & head on frozen features & $0.68$ & $0.9$ \\
$5$  & new class & $0.79\pm0.04$ & $1.0$ \\
$11$ & new class & $\mathbf{0.85\pm0.03}$ & $1.1$ \\
$68^{\dagger}$ & supervised class & $0.96$ & $1.0$ \\
\hline
\multicolumn{4}{l}{\emph{classical novae}} \\
none & blind \texttt{energy} & $0.12\pm0.03$ & $1.4$ \\
$11$ & combiner over scorers & $0.35\pm0.08$ & $0.9$ \\
$11$ & head on frozen features & $0.82\pm0.07$ & $0.6$ \\
$11$ & new class & $\mathbf{0.83\pm0.05}$ & $0.8$ \\
$37^{\dagger}$ & supervised class & $0.87$ & $1.0$ \\
\hline\hline
\end{tabular}
\end{table}

\section{Reading a single report}
\label{app:diagnostics}

A classical nova hidden in the test stream shows why the four numbers of Section~\ref{sec:conformal} are reported together: its most probable class is SN\,Ia, but with a $p$-value below $\alpha$, so that the prediction set excludes the class that the classifier reports, and its credibility, $0.14$, is about a quarter of the median credibility of the SN\,Ia in the test split ($0.50$). Empty sets at looser levels mark objects that match nothing well, although the set size alone is a weak anomaly score (Section~\ref{sec:ceiling}). The probabilities of the trees are already well calibrated (top-label expected calibration error $0.0093$). As novelty scores, the width of the Venn--Abers interval of the predicted class \citep{vovk2014venn} and the sets of a split class-conditional density-ratio construction \citep{guan2022bcops}, calibrated on an independent half of the validation objects, reach areas under the curve of $0.82$ and $0.59$ on the $53$ out-of-taxonomy objects and detect none of the sixteen objects without a parent class at a nominal $1\%$ false-alarm threshold.

\bibliographystyle{aasjournal}
\bibliography{references}

\end{document}